\documentstyle[aps,prb,epsfig,multicol]{revtex}
\makeatletter
\begin{document}
\title{The Parallel Magnetoconductance of Interacting
Electrons \\ in a Two Dimensional Disordered System}
\author{Richard Berkovits$^{1}$ and Jan W. Kantelhardt$^{1,2}$
\footnote{Present address: Center for Polymer Studies and Department of Physics, 
Boston University, Boston, MA 02215, USA.}}
\address{$^{1}$ Minerva Center and Department of Physics,
Bar-Ilan University, Ramat-Gan 52900, Israel}
\address{$^{2}$ Institut f\"ur Theoretische Physik III, 
Universit\"at Giessen, D-35392 Giessen, Germany}
%\date{October 8th, 2001, version 2.3}
\draft
\maketitle
\begin{multicols}{2}[%
\begin{abstract} The transport properties of interacting electrons for which the spin
degree of freedom is taken into account
are numerically studied for small two dimensional diffusive clusters.
On-site electron-electron interactions tend to delocalize the
electrons, while long-range interactions enhance localization.
On careful examination of  the transport properties, we reach the conclusion 
that it does not show a 
two dimensional metal insulator transition driven by interactions.
A parallel magnetic field leads to enhanced resistivity, which
saturates once the electrons become fully spin polarized.
The strength of the magnetic field for which the resistivity saturates
decreases as electron density goes down. Thus, the numerical calculations
capture some of the features seen in recent experimental measurements
of parallel magnetoconductance.
\end{abstract}
]

There has been
much recent interest in the influence of electron-electron 
interaction ({\em eei})
on the localization properties of electrons
in  two dimensional disordered systems. Behind this renewed interest in the
topic are
new experimental observations pertaining to
the behavior of the conductance of low
density two dimensional electrons.
The conductance exhibits a crossover 
from an insulating like temperature dependence
at low densities to a metallic one at higher
densities.\cite{review01,kravchenko94}
A second transition back to an 
insulating dependence at even higher densities was also
observed.\cite{hamilton99}
This transition, which is known as the 
2DMIT (two dimensional
``metal-insulator'' transition),
has drawn a flurry of theoretical activity since it 
is at odd with the prevailing single parameter
scaling theory of localization.\cite{abrahams79}
This scaling theory asserts that 
for non-interacting electrons
all states in 2D are localized by any amount of disorder.
Since in low density systems the ratio between
the typical interaction energy and the Fermi energy, $r_s$,
is large (i.e., $r_s > 1$),
a natural explanation for the 2DMIT 
is that it is the result of the {\em eei} not taken into account
in the original scaling theory. This has
prompted an intensive theoretical effort including
analytical \cite{dobrosavljevic97} and numerical 
\cite{pikus94,talamantes96,cuvas99,benenti99,waintal99,shepelyansky99,denteneer99,vojta00}
work which tried to explain the 2DMIT as a result of delocalization
by the {\em eei}.
On the other hand, one may argue that the 
observed temperature dependence of the conductance
is not a result of a metallic zero temperature phase
but rather a manifestation of essentially
"high" temperature physics. Thus, 
some other physical mechanism (such as traps,\cite{altshuler99}
interband spin dependent scattering,\cite{papadakis99,yaish00}
temperature dependent screening\cite{klapwijk99,dassarma99,zala01}
or percolation\cite{meir99})
sets a very low temperature
scale and the observed metallic behavior occurs at higher
temperatures. Accordingly there is no 
2DMIT and the systems are insulating at zero 
temperature. This viewpoint may find
support in some recent experimental results 
which show a suppression of the 2DMIT once interband scattering 
is reduced,\cite{papadakis99,yaish00}  and from the observation that the
bulk of the metallic behavior occurs at temperatures in
which there is no quantum interference contributions to the conductance,
while weak localization corrections are observed at very low
temperatures.\cite{simmons00,senz00,altshuler01,proskuryakov01}

Since it is extremely difficult to go beyond the perturbative treatment
of strong {\em eei} in disordered systems, many numerical studies have
been performed in order to clarify the role played  by  the {\em eei} 
on the localization properties of disordered systems. While for spinless
electrons the {\em eei} qualitatively changes properties of the system
such as 
many particle energy level
statistics,\cite{berkovits94,talamantes96,cuvas99,shepelyansky99}
persistent current flow patterns \cite{berkovits98,benenti99,selva00a,benenti00}
and charge density response to an external
perturbation,\cite{waintal99}
it does not lead to delocalization of electrons at the Fermi level, 
which would manifest
itself in enhanced conductivity.\cite{berkovits96,vojta98,berkovits01}

The electron spin degree of freedom nevertheless plays an important role 
in the so called 2DMIT. When the influence of the {\em eei}
on the conductance of disordered systems
is considered using a combination 
of perturbative and renormalization group techniques \cite{finkelstein84}
it leads to the conclusion that there is a divergence in the Cooperon channel at
medium interactions, while the ladder channel monotonously decreases
as function of the interaction strength. Thus, for polarized (spinless)
electrons {\em eei} should indeed decrease the conductance at zero temperature,
while the situation for unpolarized electrons is unclear.

The purpose of this paper is to investigate the influence of the 
{\em eei} on the transport through a disordered system once the spin degree
of freedom is taken into account. In previous studies in which 
spin was considered the effect of 
{\em eei} on various properties of mesoscopic systems, such as the persistent 
current \cite{ramin95,kamal95,romer95,wang96,kotlyar01,selva00} and the 
addition spectrum and spin polarization of quantum dots in the
Coulomb blockade regime
\cite{prus96,berkovits98a,brouwer99,eisenberg99,baranger00,jacquod00,kurland01,benenti01} was 
studied. 
In this paper we shall study 
the role played by {\em eei} on the tunneling amplitude. 
By evaluating the statistical 
properties of the tunneling amplitude of an interacting system one obtains 
information on 
the localization of the electrons, akin to the properties 
revealed by the statistics of 
the wave functions of non-interacting electrons, 
which is directly connected to the 
conductance.\cite{berkovits01}

The effect of {\em eei} on the localization of the 
system is quite subtle. While
the long-range part of the interaction always 
enhances the localization of the electrons, 
the short-range (Hubbard) part tends to delocalize the electrons.
Thus, when both interactions are taken into account there is a competing
influence, resulting in some delocalization for weak interactions while
the localization is enhanced for stronger interactions.
Although there are some superficial similarities to the 2DMIT, 
there are nevertheless some important 
differences, which in our opinion rule out delocalization due to 
the Hubbard 
interaction as an explanation for the 2DMIT. 
In order for a substantial delocalization to occur the Hubbard interaction
must be significantly stronger than the long-range Coulomb part, i.e.,
the system is in the high density regime where
good screening occurs. Indeed, for any reasonable estimation of interaction
parameters,
this delocalization occurs for 
electron densities corresponding to $r_s < 2$, while the 2DMIT 
occurs at much lower 
densities corresponding to  
$r_s \sim 4-20$. 

Another striking feature exhibited by some systems for which a 2DMIT 
was observed is the strong decrease in the conductance once an in-plain
magnetic field is
applied.\cite{simonian97,pudalov97,simmons98,yoon00,shashkin01,sarachik01}
This decrease coincides with the appearance
of magnetization \cite{okamoto99,vitkalov00}
and saturates around the point for which 
the system becomes fully spin polarized.
The saturation field decreases as the density goes down and
becomes zero at the ``metal-insulator'' transition point.\cite{shashkin01,sarachik01}
Naturally, one would like to relate
the decrease in conductance to
the appearance of spin polarization.
In fact, we find that 
the tunneling amplitude depends 
on the ground-state spin state once Hubbard interactions are considered, and
since the application of an external 
in-plain magnetic field changes the ground-state spin \cite{kurland01} 
it leads to a dependence of the conductance on the spin polarization.
We observe 
a positive magnetoresistance, which is akin in some aspects to the
experimental one, as well as saturation of the resistance at a critical
magnetic field corresponding to full polarization.
Unlike the enhancement in the conductance due to the Hubbard interaction
this behavior is robust to the addition of long-range Coulomb interactions.
Moreover, the critical magnetic field is reduced as the interaction strength
is enhanced (i.e., corresponding to lower densities). Nevertheless, 
as we shall discuss later, a direct comparison between
the numerical and experimental behavior leaves some open questions.

We consider the following  
tight-binding Hamiltonian:
\begin{eqnarray}
{\hat H} = \sum_{k,j;\sigma} \epsilon_{k,j} n_{k,j;\sigma}
- g \mu_{\rm B} B \hat S_z  \nonumber \\
- V \sum_{k,j;\sigma} [a_{k,j+1;\sigma}^{\dag} a_{k,j;\sigma} + 
a_{k+1,j;\sigma}^{\dag} a_{k,j;\sigma} + h.c. ] \nonumber \\ 
+ U_{\rm H} \sum_{k,j} n_{k,j;+{1 \over 2}} n_{k,j;-{1 \over 2}} 
\nonumber \\ + U \sum_{k,j>l,p;\sigma,\sigma'} (n_{k,j;\sigma} - K)
(n_{l,p,\sigma'} - K) s / |\vec r_{k,j} - \vec r_{l,p}| ,
\label{hamil}
\end{eqnarray}
where $\vec r=(k,j)$ denotes a lattice site, $a_{k,j;\sigma}^{\dag}$ 
is an electron creation operator (with spin $\sigma=-{1 \over 2},
+{1 \over 2}$), the number operator is 
$n_{k,j;\sigma}=a_{k,j;\sigma}^{\dag} a_{k,j;\sigma}$,
$\epsilon_{k,j}$ is the site energy, chosen randomly between $-W/2$ and 
$W/2$ with uniform probability, $V=1$ is a constant hopping matrix 
element, $K=\nu$ is a positive background charge equal to the electronic filling and
$s$ is the lattice constant.
As discussed above, there are two types of electron-electron 
interaction.  The Hubbard interaction $U_{\rm H}$ due to the repulsion 
of electrons with opposite spin on the same site has a short range
only, while the Coulomb interaction $U$ has a long range.  We include
an in-plain magnetic field $B$, that couples only to the total 
$z$-component of the electron spin $\hat S_z$.

We consider systems composed of $N=4$ electrons residing on 
$6 \times 6$ lattices and systems of $N=6$ electrons on $4 \times 4$
lattices, corresponding to fillings of $\nu = 1/9$ and $3/8$, 
respectively.  We chose $W=5$ and $8$, respectively, so that the 
single particle localization length is comparable to the system 
sizes.  Hard wall boundary conditions are chosen, since the 
lowest single-electron state for periodic boundary conditions 
is actually much more delocalized than neighboring states, resulting in
an unusual behavior of the conductance close to the fully polarized
state.\cite{berkovits01a}  It is of course not
possible here to directly mimic the experimental procedure in which 
electron density is varied.  Instead, the physical content of this 
density variation can be captured by controlling the ratio of the Fermi 
energy to the interaction energy.  In the present model, it is achieved 
simply by changing the interaction strengths $U$ and $U_{\rm H}$ while 
keeping other parameters constant. The value of $U$ may be related to the
electronic density via $U=V\sqrt{4 \pi \nu} r_s$. On the other hand, 
there is no generally accepted 
value for the ratio between $U_{\rm H}$ and $U$ for 2DEG.
In Hubbard's original work the ratio was estimated
as $U_{\rm H}=(10/3)U$ for weakly overlapping hydrogen
like wave-functions,\cite{Hubbard} but there is no estimation of the
ratio for relevant Si or GaAs samples.
We therefore will investigate both physical limits, $U/U_{\rm H} = 1$
and $U/U_{\rm H} = 0$, as well as other intermediate values.

We carry out our exact diagonalization in the subspace of the total
number of electrons $N$ and the total spin component $S_z = M-N/2$
with $M$ being the number of spin up electrons.  Since there is no
mechanism for spin flip in the model, the many-particle wave functions 
with different values of $S_z$ do not interact, and 
they can be calculated separately by block diagonalization.
Using the Lanczos method we obtain the many-particle eigenvalues 
$\varepsilon_\alpha^{N,S_z}$ and eigenfunctions $|\alpha^{N,S_z}\rangle$.
Because of spin symmetry, as long as $B=0$, 
$\varepsilon_\alpha^{N,-S_z}=\varepsilon_\alpha^{N,S_z}$ and 
$|\alpha^{N,-S_z}\rangle=|\alpha^{N,S_z}\rangle$.

The zero temperature local tunneling amplitude $\langle 0^{N} |
a_{\vec r,\sigma}^{\dag} |0^{N-1}\rangle$ between the ground state of $N$ and
$N-1$ electrons can be employed here in order to characterize the
transport properties of the many-particle interacting system.  It has
the advantage that only the ground state energy and eigenvector for $N$
and $N-1$ electrons need to be calculated.  The use of the tunneling 
amplitude in this context has been motivated and substantiated in our 
previous work.\cite{berkovits01,Kantelh01}  If one compares the
tunneling density of states (TDOS) $\nu(\varepsilon)$ in the independent 
particle approximation on the one hand and for the many-body interacting 
system on the other hand, it becomes evident that the tunneling 
amplitude $\langle 0^{N} |a_{\vec r,\sigma}^{\dag} |0^{N-1}\rangle$ replaces 
the single electron wave function.  The same is true for the 
transmission $t(\vec r,\vec r ~',\varepsilon,\sigma)$ of an electron with 
energy $\varepsilon$ and spin $\sigma$
between two points $\vec r$, $\vec r~'$ on the 
interface of the system with external leads, which is related to 
the conductance $\sigma(\varepsilon)$ through the Landauer formula 
\cite{landauer57} $\sigma(\varepsilon)= (e^2/h) \sum_{\vec r, \vec r~',\sigma} 
| t(\vec r,\vec r ~',\varepsilon,\sigma) |^2 $, 
where the sum $\vec r,\vec r~'$ is over all points
on the interface.  This behavior suggests that the tunneling amplitude is
the appropriate quantity to replace the single electron wave function
in studying transport properties of interacting systems.  A similar 
procedure is employed in Ref. \cite{jeon99} in order to generalize the 
concept of inverse participation ratio for interacting systems.

Note, however, that once interactions are present a many particle state 
is a superposition of many different Slater determinants.  Hence,
the tunneling amplitude is not normalized, as is the single electron 
wave function, which is the result of the fact that the spectral weight for
interacting systems is not necessarily equal to one.
Therefore, in order to study the 
influence of {\em eei} on quantum localization it is useful to define 
an effective tunneling amplitude $\phi (\vec r,\sigma) = \langle 0^{N} | 
a_{\vec r,\sigma}^{\dag} |0^{N-1}\rangle / (\sum_{\vec r} \langle 0^{N} | 
a_{\vec r,\sigma}^{\dag} |0^{N-1}\rangle^2)^{1/2}$.
In the following, the effective tunneling amplitude $\phi(\vec r,\sigma)$ 
is traced for several Hubbard and Coulomb interaction strengths 
$U_{\rm H}$ and $U$ as well as for different in-plain magnetic fields 
$B$.  In order to analyze the tunneling amplitude and to derive the 
degree of localization, we calculate the inverse participation ratio 
$P = \sum_{\vec r, \sigma} \vert\phi(\vec r)\vert^4$ averaged over $100$ 
realizations.  

It is also possible to discuss the tunneling amplitudes between the lowest
eigenvectors of a given spin sector $S_z$.
Of course, when adding an additional electron, the 
total spin component $S_z$ can either increase or decrease by one half.  
Denoting the lowest eigenvalue for a given spin sector $S_z$ with $N$ 
electrons as
$|0^{N,S_z}\rangle$, the tunneling amplitudes between the lowest
eigenvectors for a given spin sector
is $\langle 0^{N,S_z^{\rm final}} | 
a_{\vec r; \pm 1/2}^{\dag} |0^{N-1,S_z^{\rm initial}}\rangle$
with $S_z^{\rm final} = S_z^{\rm initial} \pm 1/2$ for all values of 
$S_z^{\rm initial}$, i.e. $S_z^{\rm initial} = {1 \over 2}, 
{3 \over 2}$ for $N=4$ and $S_z^{\rm initial} = {1 \over 2}, 
{3 \over 2}, {5 \over 2}$ for $N=6$.  The results for the corresponding 
participation ratios are show in Fig.~1 as a function of $U_{\rm H}$ 
without Coulomb interaction ($U=0$).  While the tunneling amplitudes 
of all possible spin channels have very similar participation ratios 
for $U_{\rm H}=0$, some of them are significantly increasing 
for larger Hubbard interaction, indicating weaker localization.  In the 
case of completely polarized electrons, $S_z = {3 \over 2} \to 2$ in (a)
and $S_z = {5 \over 2} \to 3$ in (b), the Hubbard interaction has no 
effect, since it couples electrons of opposite spin only.  It is interesting to
note that there seems to be two distinct magnitudes of enhancement.
The inverse participation ratio for transitions  in which 
$S_z^{\rm final} = S_z^{\rm initial} - 1/2$ 
are substantially larger than for transitions where
 $S_z^{\rm final} = S_z^{\rm initial} + 1/2$.
This may be the result of the fact that for transitions with 
$S_z^{\rm final} = S_z^{\rm initial} - 1/2$ the additional spin-down 
electron can join many spin-up "partners", thus enhancing the effectiveness 
of the Hubbard interaction.  For transitions with 
$S_z^{\rm final} = S_z^{\rm initial} + 1/2$ the influence of the Hubbard 
interaction is diminished, because the relative number of possible pairs 
of electrons with opposite spin, that can share sites, is reduced.
One may also observe that $\langle P^{-1} \rangle$ for the $4 \times 4$
lattice shows some decrease for 
$U_{\rm H} \ge 5 V$, while for the $6 \times 6$
lattice $\langle P^{-1} \rangle$ more or less saturates. We attribute this
behavior to the fact that for the smaller lattice the electronic density
is higher ($\nu=3/8$) resulting in an observable effect
of the Mott-Hubbard transition, while for the lower density ($\nu=1/9$)
pertaining to the larger lattice the Mott-Hubbard transition is less
pronounced.

In order to find out which of the different possible spin channels is 
relevant for the zero temperature transport, the energies 
$\varepsilon_\alpha^{N-1,S_z^{\rm initial}}$ and 
$\varepsilon_\alpha^{N,S_z^{\rm final}}$ of the corresponding 
multi-particle eigenstates have to be investigated.  
$S_z^{\rm initial}$ and $S_z^{\rm final}$ have to be selected such,
that both energies have their minimal value for each configuration.
With no magnetic field, for a ground state spin value $S$, all
energies $\varepsilon_0^{N-1,S_z=-S,-S+1, \ldots, S-1,S}$
are degenerate. Once an infinitesimal magnetic field $B \rightarrow 0^+$
is present (which we assume here),the ground state corresponds to the 
maximal value of $S_z=S$.
Only if the condition $S_z^{\rm final} = S_z^{\rm initial} \pm 1/2$ 
can be fulfilled after this minimizing procedure, the corresponding
configuration has nonzero tunneling amplitude at zero temperature. 

Figure 2 shows the average values of the initial and final spins
(open symbols, right scale) for zero temperature transport as well as
the average values of the participation ratio of the corresponding
tunneling amplitudes (filled symbols, left scale) as function of the
Hubbard interaction $U_{\rm H}$, assuming $B \rightarrow 0^+$.  
For the participation ratios, three
different averaging procedures are compared:
arithmetic average $\langle P^{-1} \rangle$,
typical average $\exp - \langle \ln P \rangle$, and
geometrical average $1 / \langle P \rangle$. 
They give slightly different values, since the participation ratios
are strongly fluctuating.  But all of them for both considered
systems show the same qualitative behavior:  Upon increasing Hubbard
interaction, the participation ratios for the zero-temperature transport
are increasing as long as $U_{\rm H} \le 5 V$, and they become weakly dependent
on $U_{\rm H}$ for larger interactions.  Hence, the zero-temperature
transport in the 2DEG is enhanced by the Hubbard interaction.  The
enhancement is rather weak, though, reaching about 50\% for the
$6 \times 6$ lattice and 70\% for the $4 \times 4$ lattice.  The range
of $U_{\rm H}$ where the enhancement occurs, cannot be directly
compared to the experiments, since the relevant parameter $r_{\rm s}$
is rather related to the Coulomb interaction $U$ (which is zero here)
than to the Hubbard interaction $U_{\rm H}$. 

Next, we want to investigate, how the zero-temperature transport is
affected by an in-plain magnetic field.  While there would be some
modification of the electrons' orbits for perpendicular magnetic fields
(see e.g. \cite{Kantelh01}) the parallel magnetic field interacts only
with the electrons' spins.  Since the multi-particle eigenfunctions
$|\alpha^{N,S_z}\rangle$ have already been calculated as eigenfunctions
of the spin operator $\hat S_z$, their energy levels
$\varepsilon_\alpha^{N,S_z}$ are just shifted to
$\varepsilon_\alpha^{N,S_z} - g \mu_{\rm B} B S_z$ according to
Eq.~(\ref{hamil}).  The eigenfunctions $|\alpha^{N,S_z}\rangle$ and
the corresponding tunneling amplitudes are not changes, but with
$B \ne 0$ a different eigenstate might become the ground state. 
With increasing $B$, a higher degree of polarization corresponding
to larger $S_z$ becomes favorable.\cite{kurland01}
Figures 3(a) and 3(b) show the average values of the ground state
spins $S_z^{\rm initial}$ (for the 3 electron system) and
$S_z^{\rm final}$ (for the 4 electron system) as function of the
in-plain magnetic field $B$ for 4 values of $U_{\rm H}$.  In both
figures, a monotonously increasing polarization is observed, as
expected. 

Now, zero-temperature transport can be traced as function of $B$. 
One just has to average the participation ratios of the tunneling
amplitudes for the corresponding ground states, provided the
condition $S_z^{\rm final} = S_z^{\rm initial} \pm 1/2$ is fulfilled. 
If this condition fails, the tunneling amplitude is zero for the
corresponding configuration and $B$.  The results of this $B$-dependent
averaging procedure are shown in Fig.~3(c) for $3 \to 4$ electrons
on a $6 \times 6$ lattice (upper part) and $5 \to 6$ electrons on a
$4 \times 4$ lattice (lower part), again for 4 values of $U_{\rm H}$.
In the non-interacting case there is a weak dependence of the
conductance on the magnetic field as a result of the weak dependence of
$\langle P^{-1} \rangle$ on the single particle state seen in Fig. 1 
for $U_{\rm H}=0$.
For nonzero Hubbard interaction $U_{\rm H}$ we observe a drop in 
$\langle P^{-1} \rangle$
 in the range of $g \mu_{\rm B} B < \Delta$, where $\Delta$ is the mean single 
particle level spacing.  This behavior is the result of the larger participation 
ratio exhibited by the $S_z^{\rm final} = S_z^{\rm initial} - 1/2$ transition 
compared to the $S_z^{\rm final} = S_z^{\rm initial} + 1/2$ transition.  As can 
be seen in Fig. 3, when $g \mu_{\rm B} B < \Delta /2$, for all realizations, the 
ground state corresponds to $S_z^{\rm initial}=1/2$ for the odd number of 
electrons, while more and more realizations change from $S_z^{\rm final}=0$ 
to $S_z^{\rm final}=1$ as $B$ increases.  This leads to the decrease of the 
participation ratio of the average tunneling amplitude, corresponding to a 
decrease in the conductance.  The initial decrease is followed by intermediate 
maxima and minima, since, as $B$ increases, the average $S_z$ increases leading 
to a local peaks in the conductance each time that the $S_z^{\rm final} = 
S_z^{\rm initial} - 1/2$ becomes more prevalent.  Once a larger portion of 
realizations become fully polarized (for large magnetic fields $B$), the 
conductance
drops to the same saturated value for all $U_{\rm H}$ values.  The values 
$B_{\rm sat}$ of the magnetic field for which the systems become fully polarized 
are shown in Fig. 4(a) for both system sizes.  It can be seen, that $B_{\rm sat}$ 
decreases with increasing Hubbard interaction $U_{\rm H}$, but it never vanishes.

Our findings somewhat resemble the observations of recent 
experiments:\cite{review01}  The conductivity of 2DEG semiconductor
devices, that seem to show a "metallic phase", is reduced with increasing
in-plain magnetic field, until a constant conductivity is reached
for $B > B_{\rm sat}$.  Although this qualitative behavior is strikingly
similar to the results of our calculation, the magnitude of the
decrease we observe is much smaller than that in the experiments. 
For silicon devices decreases by up to two orders of magnitude have
been observed depending on the temperature,\cite{simonian97,pudalov97}
while for GaAs devices, decreases by a factor of 3 
have been reported.\cite{simmons98,yoon00}
Another problem in relating our results to the experiments is 
the dependence of the saturation field $B_{\rm sat}$ on 
$U_{\rm H}$. In the experiments 
$B_{\rm sat} \propto (\nu-\nu_c)^{\delta}$, where
$\nu_c$ is the density of the 2DMIT and $\delta=1$ 
for a wide range of densities,\cite{shashkin01}
while $\delta \approx 0.6$ close to $\nu_c$ elsewhere.\cite{sarachik01}
Since we do not identify a metal-insulator transition in
our model it is not clear how to directly relate the above
experimental observations to our data. Moreover,
the experimentally relevant parameter is rather  
related to the Coulomb interaction $U$ (via $r_{\rm s}$)
(which is zero here) than to the Hubbard interaction $U_{\rm H}$.

Nevertheless, it is possible to relate $B_{\rm sat}$ to $U_{\rm H}$
using the following consideration:
The lowest average energy of
the Hamiltonian given in Eq.~\ref{hamil} at a given spin $S$
may be approximated by:\cite{kurland01}
\begin{eqnarray}
E(S) - E(0) = \Delta S^2 - JS(S+1) - g \mu_{\rm B} B S,
\label{jeq}
\end{eqnarray}
where $J$ is the averaged exchange energy which depends on
$U$ and $U_{\rm H}$.  Figure 4(b) shows the values of $J$ determined
from the eigenvalues $E(S_{\rm max})$ and $E(0)$ of the systems via 
Eq.~(\ref{jeq}) for $B=0$ and $U=0$.  A power-law relation $J \propto 
U_{\rm H}^{\alpha}$ with an exponent $\alpha$ in the range from 0.25 
to 0.5 may be deduced from our data, and this relation is consistent 
with results obtained in a previous study of the disordered Hubbard 
model.\cite{folk01} 
The saturation field $B_{\rm sat}$ corresponds to the field in which the
energy becomes minimal for complete polarization, i.e. $\partial E(S) /
\partial S = 0$ with $S=S_{\rm max}$, resulting in
\begin{eqnarray}
g \mu_{\rm B}B_{\rm sat} = 2 \Delta S_{\rm max} - J(2S_{\rm max}+1) .
\label{bsat}
\end{eqnarray}
For small values of $S$ it is better to replace the
derivative leading to Eq.~(\ref{bsat}) by the discrete difference leading to
$g \mu_{\rm B}B_{\rm sat} = \Delta (2S_{\rm max}-1) - 2 J S_{\rm max}$.
In Figure 4(b) the value of the averaged exchange energy $J$ obtained 
from $B_{\rm sat}$ via this equation is compared to the $J$ obtained
from the eigenvalues.  The Figure shows that the relations work reasonably 
well in describing $B_{\rm sat}$. 
Since $J$ increases as function of $U_{\rm H}$ the 
saturation field decreases.  
Thus, the same physics
of complete spin polarization at the saturation field describes
both the experiment\cite{dolgopolov00,shashkin01} and the numerical results.
It is important to note though, that since the exchange\cite{kurland01} 
saturates at values lower than $\Delta$, $J(U_{\rm H}
\rightarrow \infty) \sim 0.6 \Delta$, 
there will be no ferromagnetic transition for any value of
$U_{\rm H}$. This is clear from Fig.~2 where the average value of the 
spin saturates at low values. Therefore $B_{\rm sat} \ne 0$ for any value of
$U_{\rm H}$, and a model with no long-range interactions can not describe 
the experimentally observed $B_{\rm sat} \propto (\nu-\nu_c)^{\delta}$.

Thus, it is clear that one must move beyond the short-range Hubbard 
interaction and include a long-range Coulomb interaction in order to
present a realistic picture of the 2DMIT behavior. As discussed above
there is no established relation between $U$ and $U_{\rm H}$, so we
shall begin by presenting the highest possible ratio between them,
i.e., $U=U_{\rm H}$. The participation ratio as function of $U$ is
presented in Fig. 5(a). An initial enhancement of $P^{-1}$, which peaks
at $U=2V$ is seen, while for larger values of interaction a steady decrease
is observed. This behavior is attributed to the interplay between the
short and long-range components of the {\it eei}. Short-range interactions
enhance the participation ratio of the tunneling amplitudes,
while the long-range interactions suppress it. This is demonstrated in
Fig. 5(b) where we keep $U_{\rm H}=10 V$ fixed and tune the value of $U$.
A monotonous decease in $P^{-1}$ as function of $U$ is evident.
The influence of the long-range interaction on the spin polarization
can also be gleaned from Fig. 5. The short-range interactions lead to spin
polarization, while the long rang interaction tends only to 
slightly nudge the ground state spin. 

The dependence of the participation ratio on the in-plain magnetic
field in the presence of long range {\it eei} is depicted in Fig. 6.
The main feature of Fig. 3, i. e., the suppression of $P^{-1}$ by the 
magnetic field and saturation at high magnetic fields remains intact.
Thus, long range Coulomb interactions do not overturn the main
conclusions garnered from the magnetoconductance 
for short-range interactions.  
The dependence of the saturation field $B_{\rm sat}$ on the Coulomb 
interaction strength $U$ shown in Fig. 4(a) is also similar to the
dependence on $U_{\rm H}$ for $U=0$.
There are some differences though. The saturated value of
$P^{-1}$ decreases as $U$ becomes stronger, which fits our previous
observation that long-range interaction localizes the system.
The saturation field depends on the interaction strength, but 
in contrast to the Hubbard interaction we can not rule out
that $B_{\rm sat}=0$ for a finite value of $U$.
If $B_{\rm sat}=0$ indeed occurred,
the value of interaction for which the field is equal to zero 
would be much larger than the values of interaction
considered here. Therefore, we cannot compare the behavior of
$B_{\rm sat}$ to the one seen in the experiment based on the available 
data.

Once long-range interactions are taken into account one can use the
relation $U=V\sqrt{4 \pi \nu} r_s$, resulting in a connection between
interaction strength and electronic density. It would be tempting to interpret
Fig. 5(a) as a signature of a 2DMIT, but this explanation raises difficulties
which make it rather doubtful. 
The first "transition" from a more insulating behavior to a more metallic
one (see Fig. 5(a))
is influenced almost entirely by the Hubbard term as can be seen from its 
similarity to Fig. 2. At $U \approx 2V$, the point where the enhancement 
due to the short-range interaction saturates, the suppression due to long-range
interactions kicks in. Since the saturation point is determined by 
$U_{\rm H}$, while $r_s$ depends on $U$ the corresponding electronic density 
for which the 
peak appears to depend on the ratio of $U_{\rm H}/U$.
For the lowest possible ratio (i.e., $U_{\rm H}/U=1$) the peak occurs at
densities corresponding to $r_s \sim 1.6$
which is low compared to the experimentally observed region
of metallic behavior\cite{review01} corresponding to $r_s \sim 4-20$.
Similar behavior for the site occupation
number, which gives some indication on the degree of
localization in the system, has been seen by Selva and Pichard.\cite{selva00}
A higher ratio of $U_{\rm H}/U$ will result in this 
 peak appearing for even lower values of $r_s$. 
The high densities (low values of $r_s$)
for which the peak in the participation ratio appears, in conjunction with 
the fact that long-range Coulomb interactions always suppress
$P^{-1}$ (see Fig. 5(b)) lead us to the conclusion that numerical studies
of small clusters do not support the notion of an {\it eei} driven
metal insulator transition.

On the other hand, the numerical model does seem to reproduce
the positive in-plain magnetoresistance and 
the saturation of the resistance at a critical magnetic field seen
in recent
experiments.\cite{simonian97,pudalov97,simmons98,yoon00,shashkin01,sarachik01}
The physical origin of the saturation is 
spin polarization of the electron. Similar explanations were proposed
for the experimental origin of the saturation.\cite{dolgopolov00,shashkin01}
Further studies of higher values of interactions are needed to determine
whether quantitative comparison with 
the experimentally observed $B_{\rm sat} \propto (\nu-\nu_c)^{\delta}$
can be obtained. Furthermore, if such behavior was indeed observed, the origin
of $\nu_c$ would have to be clarified, since it can not be the critical density
of the 2DMIT which is not seen in our model.

Beyond relating our results to existing experimental data, one would like 
to glean some relevance for possible future experiments.  As we pointed out
in the introduction the physical situation corresponding to a short
range {\it eei} is deep in the metallic regime in which the long-range
interaction is perfectly screened.
Thus, the enhancement of the conductance in Fig. 2 is not relevant to the 
existing body of experimental data on the 2DMIT, but it might be relevant 
to an experimental double gate set-up in which the long-range part of the 
{\it eei} is screened.\cite{gefen01} According to Fig. 2 the conductance of
such a set-up should be higher than for a single or no gate set-up for 
the same 2DEG sample. 

In conclusion,
the conductance of interacting electrons for which the spin
degree of freedom is taken into account exhibits an intricate dependence
upon {\it eei} and
parallel magnetic field.
Hubbard on-site interactions enhance the conductance 
while long-range Coulomb interactions suppress it.
The interplay between the two can lead to a region of enhanced conductance, 
but this region is at densities corresponding to $r_s \approx 1$, which are 
too high to be relevant to the 2DMIT.
A parallel magnetic field reduces the conductance until it
saturates once the electrons become fully spin polarized.
The saturation field 
decreases for lower electron density. These 
features are seen in recent experimental measurements
of parallel magnetoconductance. Further investigation,
at higher values of interaction, are needed in order
to clarify whether other features may be explained by
this model. 

RB thank the Israel Science Foundations Centers of Excellence Program 
and JK thanks the Minerva Foundation and the Deutsche Akademische
Austauschdienst (DAAD) for financial support.

\end{multicols}

\begin{figure}\centering
\epsfxsize14cm\epsfbox{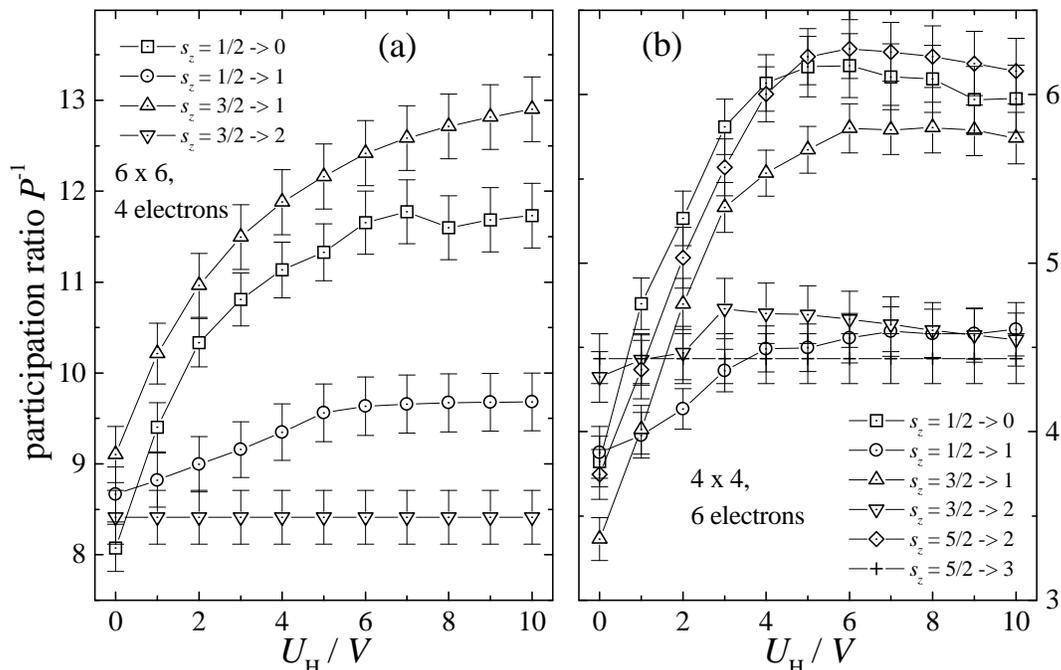}
\caption{The participation ratio $P^{-1}$ of the tunneling amplitude 
for the different spin channels is shown as function of the Hubbard 
interaction strength $U_{\rm H}$ for (a) $3 \to 4$  electrons on a 
$6 \times 6$ lattice, and (b) $5 \to 6$  electrons on a $4 \times 4$ 
lattice, both without Coulomb interaction.  The symbols correspond 
to the spin transitions: 
$S_z = {1 \over 2} \to 0$ (squares), ${1 \over 2} \to 1$ (circles), 
${3 \over 2} \to 1$ (triangles up), ${3 \over 2} \to 2$ (triangles 
down), ${5 \over 2} \to 2$ (diamonds), ${5 \over 2} \to 3$ (plus).  
We have averaged $100$ realizations of disorder,
and the error bars show the standard deviations of the averages.} 
\end{figure}

\begin{figure}\centering 
\epsfxsize14cm\epsfbox{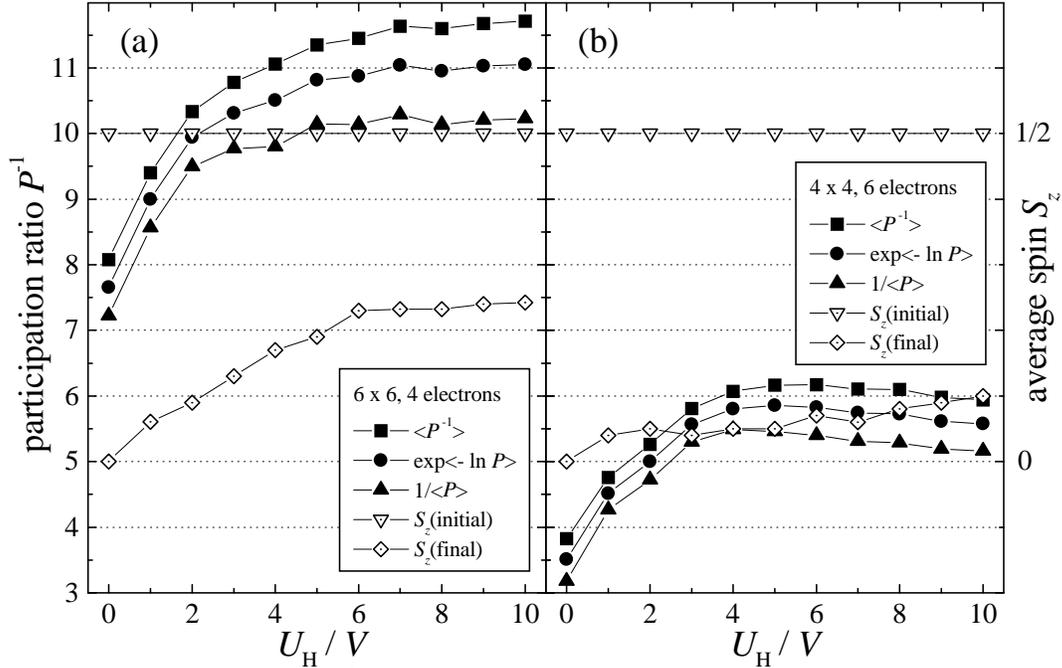}
\caption{The average participation ratio $P^{-1}$ of the zero temperature 
tunneling amplitude (filled symbols, left axes) and the average initial 
and final spin $S_z$ (open symbols, right axes) are shown as function 
of the Hubbard interaction strength $U_{\rm H}$ for (a) $3 \to 4$ 
electrons on a $6 \times 6$ lattice, and (b) $5 \to 6$ electrons on a 
$4 \times 4$ lattice, both without Coulomb interaction.  
In the averaging procedure, the ground states for both, initial and final 
state have been chosen as described in the text.  For all 100 disorder 
configurations, the tunneling amplitude is non-zero.  For the participation 
ratios, arithmetical, typical, and geometrical averages are compared.} 
\end{figure}

\begin{figure}\centering 
\epsfxsize14cm\epsfbox{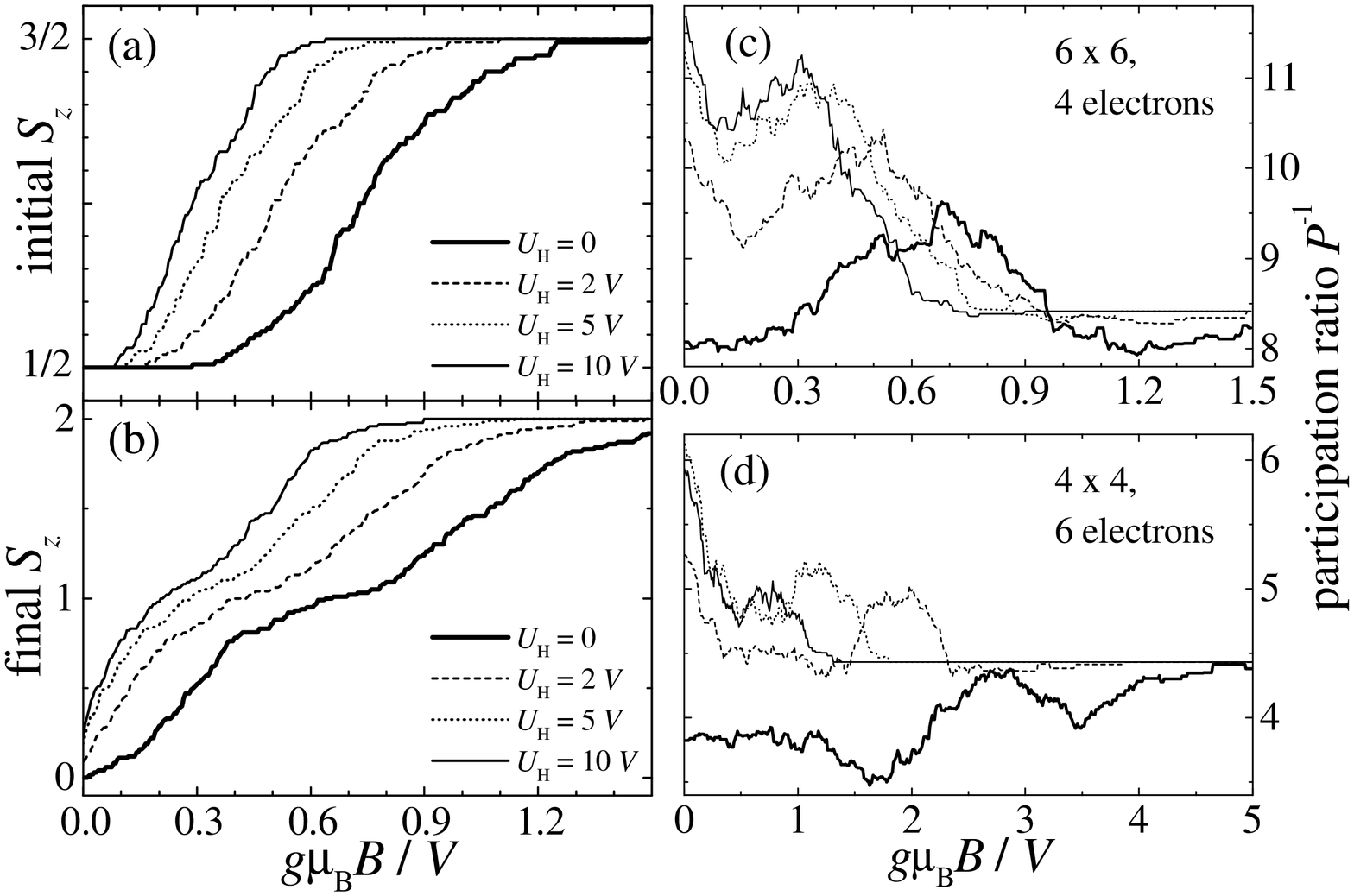}
\caption{The average (a) initial and (b) final spins $S_z$ of the ground
states are shown versus the scaled magnetic field for $3 \to 4$ electrons 
on a $6 \times 6$ lattice.  With increasing magnetic field, the ground states are
moved to higher degrees of polarization.  The different lines correspond 
to different strengths of Hubbard interaction $U_{\rm H}$ (see legends), 
while there is no Coulomb interaction.  On the right, the average participation
ratios $\langle P^{-1} \rangle$ of the zero temperature tunneling amplitude
are shown for $3 \to 4$ electrons on a $6 \times 6$ lattice (c) and $5 \to 6$
electrons on a $4 \times 4$ lattice (d).  In the (arithmetic)
averaging procedure, the ground states for both, initial and final state 
have been chosen as described in the text.  Configurations with zero 
tunneling amplitude (due to non-matching spins in initial and final ground 
states) have been disregarded in the averaging procedures.}
\end{figure}

\begin{figure}\centering 
\epsfxsize12cm\epsfbox{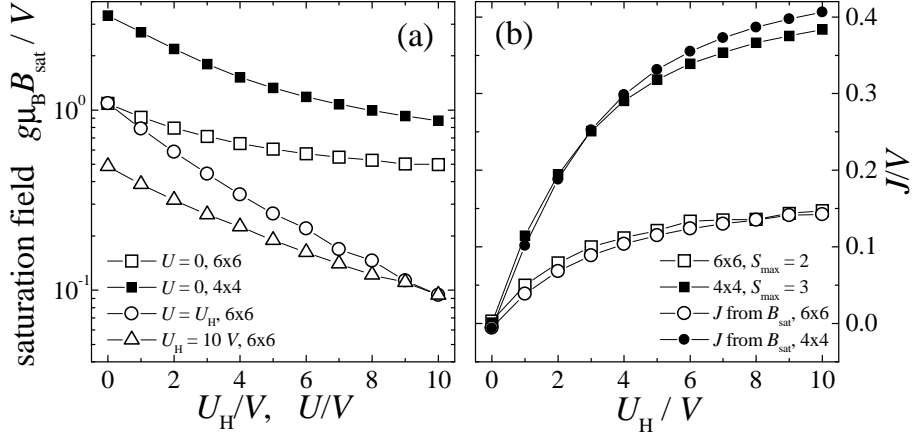}
\caption{(a) The saturation field $B_{\rm sat}$, for which the
systems become fully polarized and the average participation ratios 
$P^{-1}$ of the zero temperature tunneling amplitude reach their 
asymptotic value, are shown versus the Hubbard and the Coulomb interaction
strength (see legend).  The values of $B_{\rm sat}$ have been averaged for
100 configurations.   (b) The values of the average exchange energy
$J$ are shown versus $U_{\rm H}$ for $6 \times 6$ systems (open symbols)
as well as the $4 \times 4$ systems (filled symbols).  Two ways to calculate
$J$ are compared: (i) $J$ determined from the eigenvalues via Eq.~(2) for
$B=0$ (squares) and (ii) $J$ obtained from $B_{\rm sat}$ via the discrete
version of Eq.~(3) (circles).  The plot indicates good agreement, justifying
the approach to relate $B_{\rm sat}$ to $U_{\rm H}$.}
\end{figure}

\begin{figure}\centering 
\epsfxsize14cm\epsfbox{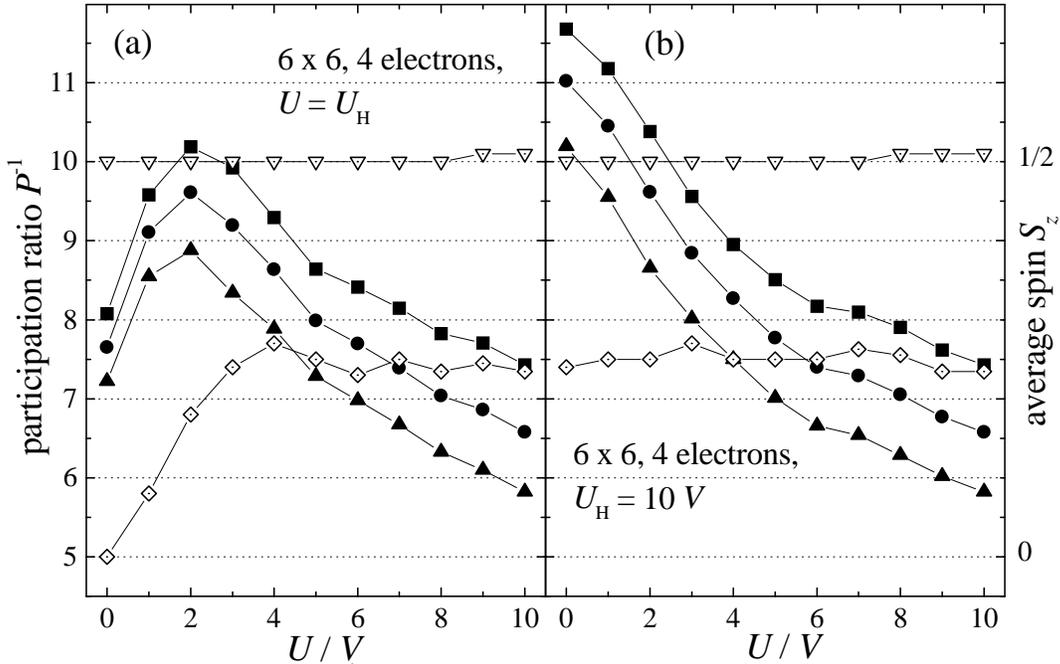}
\caption{The average participation ratio $P^{-1}$ of the zero temperature 
tunneling amplitude (filled symbols, left axes) and the average initial 
and final spin $S_z$ (open symbols, right axes) are shown as function 
of Coulomb interaction strength $U$, (a) for identical Hubbard interaction
$U_{\rm H} = U$ and (b) for fixed Hubbard interaction $U_{\rm H} = 10 V$
for $3 \to 4$ electrons on a $6 \times 6$ lattice.
The averaging procedure is the same as for Fig. 2.} 
\end{figure}

\begin{figure}\centering 
\epsfxsize14cm\epsfbox{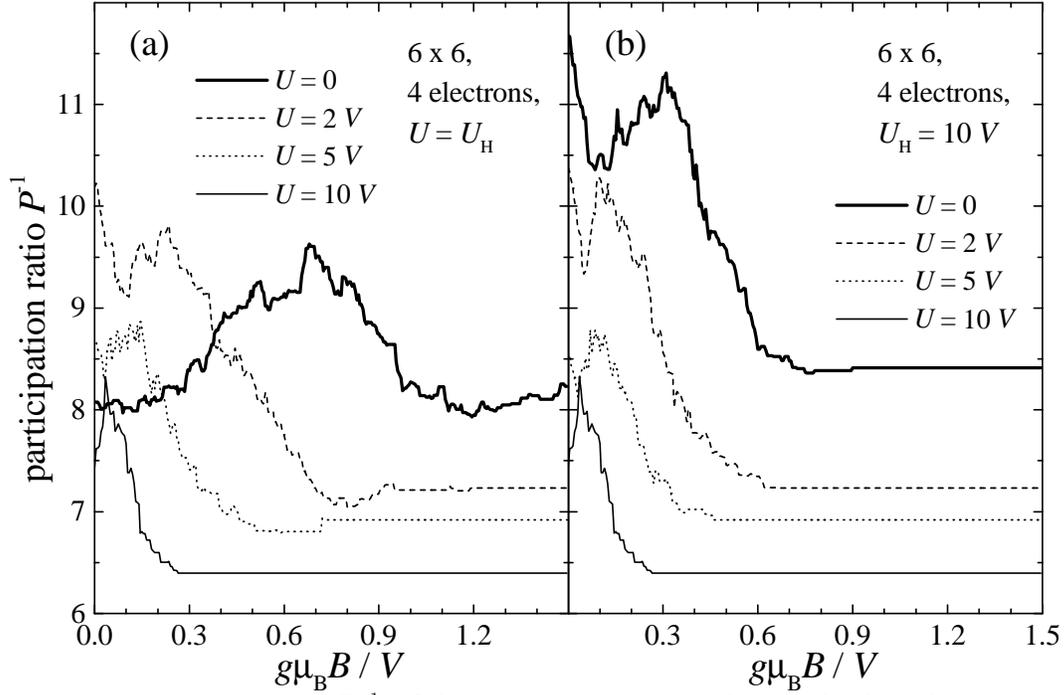}
\caption{The average participation ratio $\langle P^{-1} \rangle$ of the zero
temperature tunneling amplitude is shown versus the magnetic field $B$,
(a) for identical Hubbard interaction $U_{\rm H} = U$ and (b) for fixed 
Hubbard interaction $U_{\rm H} = 10$ for $3 \to 4$ electrons on a 
$6 \times 6$ lattice.  The averaging procedure is the same as for Fig. 3.} 
\end{figure}

\end{document}